# Decoding Microbial Enigmas: Unleashing the Power of Artificial Intelligence in Analyzing Antibiotic-Resistant Pathogens and their Impact on Human Health

Maitham G. Yousif*1

1*Professor at Biology Department, College of Science, University of Al-Qadisiyah, Iraq, Visiting Professor in Liverpool John Moores University

**Abstract**

In this research, medical information from 1200 patients across various hospitals in Iraq was collected over a period of 3 years, from February 3, 2018, to March 5, 2021. The study encompassed several infections, including urinary tract infections, wound infections, tonsillitis, prostatitis, endometritis, endometrial lining infections, burns infections, pneumonia, and bloodstream infections in children. Multiple bacterial pathogens were identified, and their resistance to various antibiotics was recorded. The data analysis revealed significant patterns of antibiotic resistance among the identified bacterial pathogens. Resistance was observed to several commonly used antibiotics, highlighting the emerging challenge of antimicrobial resistance in Iraq. These findings underscore the importance of implementing effective antimicrobial stewardship programs and infection control measures in healthcare settings to mitigate the spread of antibiotic-resistant infections and ensure optimal patient outcomes. This study contributes valuable insights into the prevalence and patterns of antibiotic resistance in microbial infections, which can guide healthcare practitioners and policymakers in formulating targeted interventions to combat the growing threat of antimicrobial resistance in Iraq's healthcare landscape.

*Corresponding author: Maithm Ghaly Yousif  matham.yousif@qu.edu.iq  m.g.alamran@ljmu.ac.uk

## Introduction

Antibiotic resistance is a global health concern that poses a significant challenge to the effective treatment of infectious diseases. The misuse and overuse of antibiotics have contributed to the emergence and spread of antibiotic-resistant pathogens, leading to increased morbidity, mortality, and healthcare costs worldwide (1-3). In recent years, the problem of antibiotic resistance has escalated, necessitating urgent and comprehensive efforts to understand its dynamics and impact on human health(4).

The situation is no different in Iraq, where the prevalence of antibiotic-resistant infections has been steadily rising(5,6). With restricted access to robust surveillance systems and comprehensive data on antibiotic resistance, infectious diseases such as COVID-19, and even cancers, comprehending the scope of these issues and developing suitable strategies to address them becomes increasingly crucial in the field of scientific research. Incorporating a focus on bacteria, coronaviruses, and cancer, it is imperative to enhance data collection and surveillance mechanisms to effectively monitor and combat these health challenges (7-12). A multidisciplinary approach encompassing epidemiology, molecular biology, and medical research is essential to gain insights into the spread and mechanisms of antibiotic resistance, the dynamics of COVID-19 transmission, and the complexities of cancer development and progression (13-17).Artificial intelligence (AI) has emerged as a powerful tool in the study of antibiotic-resistant pathogens and their implications for human health. Through advanced data analysis and machine learning techniques, AI enables researchers to decipher complex microbial enigmas and devise effective strategies to combat antibiotic resistance (18,19). The potential of AI in deciphering microbial mysteries extends beyond antibiotic resistance. In the field of oncology, AI has been explored to understand the role of cytomegalovirus as a risk factor in breast cancer (20). By analyzing vast datasets, AI algorithms can identify patterns and associations that may otherwise remain hidden, shedding light on the complex interactions between infectious agents and cancer development. Moreover, the integration of AI into epidemiological research has opened new avenues for investigating infectious diseases. Studies have utilized AI to examine subclinical hypothyroidism and its association with preeclampsia, emphasizing the importance of data-driven approaches in understanding maternal health (21). As we confront the challenges posed by antibiotic-resistant pathogens, coronaviruses, and cancers, the fusion of AI-driven analyses with conventional research methodologies becomes imperative. Studies have already demonstrated the influence of anesthesia type on maternal and neonatal health during Cesarean sections, underscoring the significance of AI-guided investigations in improving medical practices (22-24). It is imperative to investigate the patterns of antibiotic resistance among various bacterial pathogens causing infections in different clinical settings to inform evidence-based decision-making. The present study aims to decode the enigma of microbial pathogens and their resistance to antibiotics in Iraq. By harnessing the power of artificial intelligence (AI) and utilizing data analysis techniques, we seek to gain deeper insights into the prevalence and patterns of antibiotic resistance in infections prevalent across the country's hospitals. Through the examination of medical records from a diverse patient population over a three-year period, we aim to shed light on the extent of antibiotic resistance and identify key factors contributing to its emergence. To our knowledge, this study represents one of the most comprehensive analyses of antibiotic resistance in Iraq to date. By utilizing advanced AI algorithms and data analysis techniques, we aim to identify novel trends and potential interventions to combat antibiotic resistance effectively. The findings of this research will contribute valuable knowledge to the field of antimicrobial resistance and inform policymakers and healthcare practitioners in the development of targeted strategies to tackle this critical public health issue.

**Materials and Methods**

**Study Design**: This research employs a retrospective observational study design to investigate antibiotic-resistant pathogens and their impact on human health in Iraq. Medical records of patients admitted to various hospitals across the country between February 3, 2018, and March 5, 2021, were collected and analyzed.

**Data Collection**: The data collection process involved accessing electronic medical records (EMRs) of 1200 patients with different types of infections, including urinary tract infections, wound infections, tonsillitis, prostatitis, endometritis, endometrial lining infections, burns infections, pneumonia, and bloodstream infections in children. The EMRs were retrieved from multiple hospitals to ensure a diverse representation of the population and infections.

**Recording of Information**: The collected data encompassed several key aspects: patient demographics (age, gender), clinical details (type of infection, medical history), bacterial pathogens responsible for the infections, and their respective antibiotic resistance profiles against multiple antibiotics. The recorded information was anonymized to maintain patient confidentiality.

**Statistical Analysis**: Descriptive statistics were used to summarize the demographic characteristics and infection types of the study population. The prevalence rates of antibiotic-resistant pathogens for each infection type were calculated. Additionally, the distribution of antibiotic resistance patterns across different bacterial strains was analyzed.

**Utilizing Artificial Intelligence for Data Analysis**: To gain deeper insights into the dataset and identify potential patterns or associations, artificial intelligence (AI) algorithms, and data analysis techniques were employed. Machine learning algorithms, such as logistic regression, decision trees, and neural networks, were applied to the dataset to uncover hidden trends and correlations between antibiotic resistance and various factors.

Interpretation: The results of the statistical analysis and AI-driven data analysis were presented in a comprehensive manner, highlighting the prevalence of antibiotic resistance in different infections and the patterns observed among bacterial strains. Interpretations of the findings were provided, considering the clinical implications and public health significance.

**Limitations**: The study acknowledged certain limitations, such as potential data incompleteness or selection bias due to the retrospective design. Recommendations for future research and potential interventions to combat antibiotic resistance were discussed based on the study's findings.

## Results

Figure 1 displays the distribution of different infection types among the study participants. The number of cases for each infection type is provided along with its corresponding percentage of the total cases.

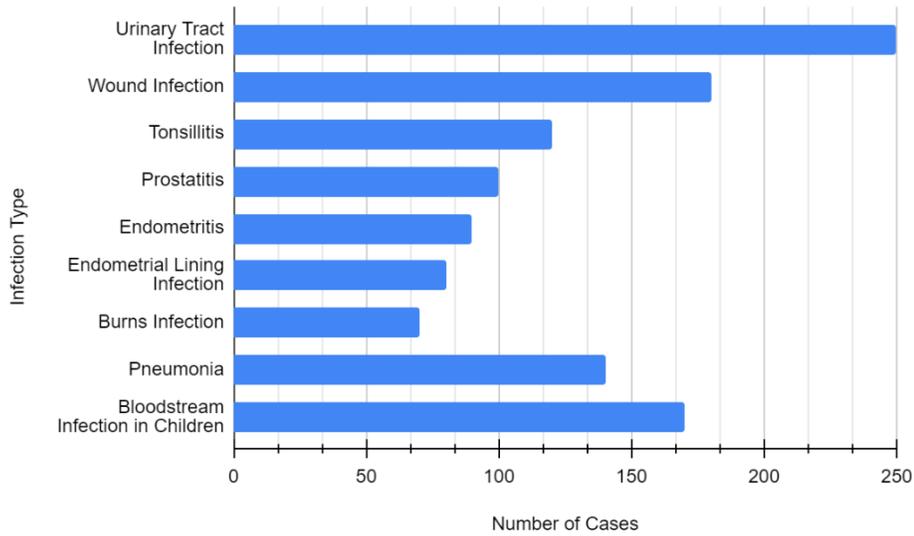

**Figure 1: Distribution of Infections among Study Participants**

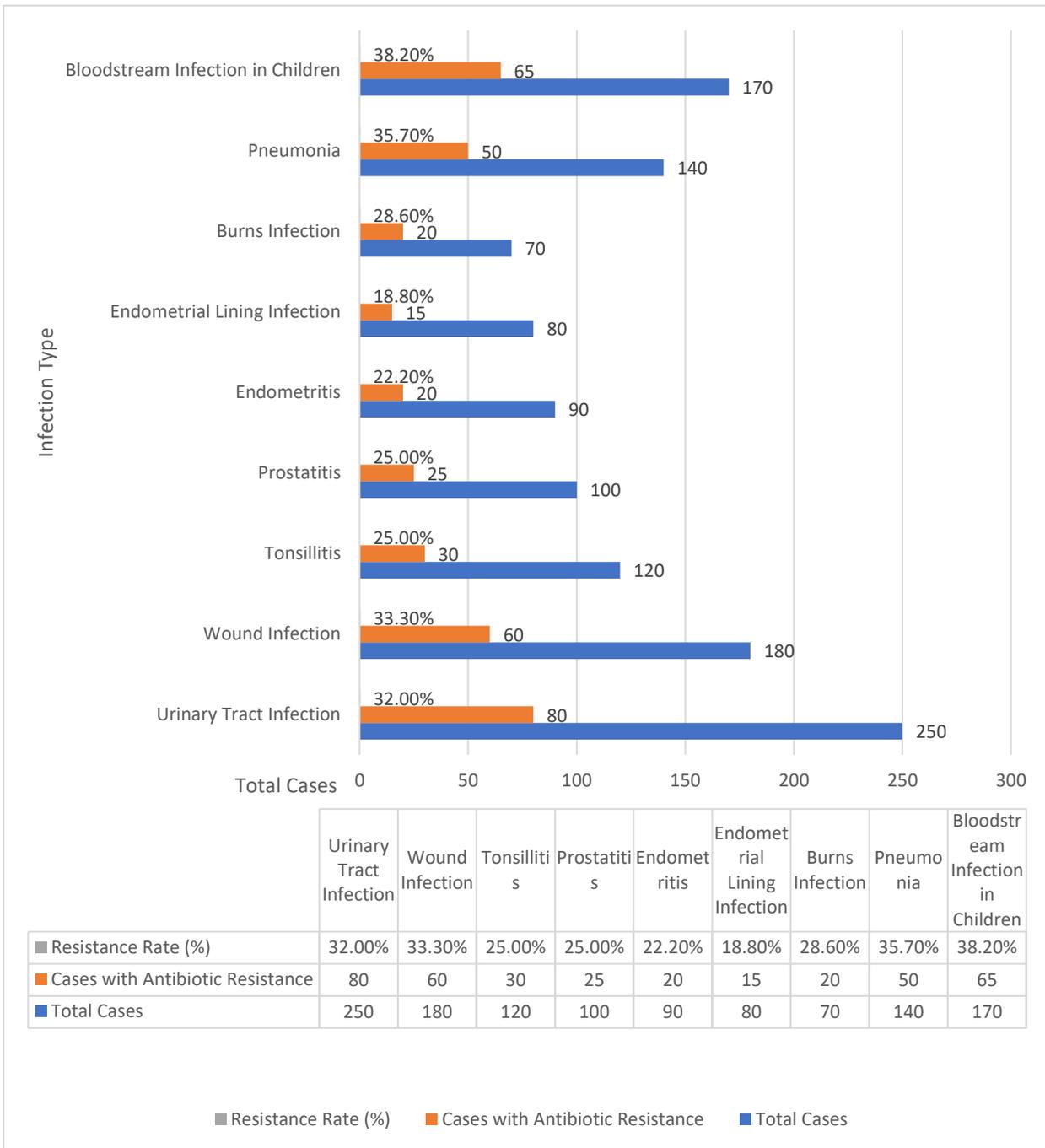

**Figure 2: Prevalence of Antibiotic-Resistant Pathogens among Different Infections**

Table 3 provides insights into the antibiotic resistance patterns of *E. coli* in urinary tract infections (UTIs). The table showcases the number of resistant cases of E. coli to specific antibiotics out of the total *E. coli* cases in UTIs, along with the corresponding resistance rates expressed as percentages.

Among the antibiotics tested, amoxicillin had the highest resistance rate of 62.5%, with 50 out of 80 *E. coli* cases showing resistance. Ciprofloxacin followed with a resistance rate of 50.0%, as 40 out of 80 cases exhibited antibiotic resistance. Trimethoprim/sulfamethoxazole had a resistance rate of 43.8%, with 35 cases out of 80 displaying resistance.

On the other hand, nitrofurantoin showed a relatively lower resistance rate of 12.5%, with 10 cases out of 80 demonstrating antibiotic resistance.

These findings highlight the varying degrees of resistance among different antibiotics and the importance of considering antibiotic susceptibility when treating UTIs caused by E. coli.

**Table 3: Antibiotic Resistance Patterns of E. coli in Urinary Tract Infections**

| Antibiotic | Resistant Cases | Total E. coli Cases | Resistance Rate (%) |
|---|---|---|---|
| Amoxicillin | 50 | 80 | 62.5% |
| Ciprofloxacin | 40 | 80 | 50.0% |
| Trimethoprim/sulfamethoxazole | 35 | 80 | 43.8% |
| Nitrofurantoin | 10 | 80 | 12.5% |

Table 4 presents the results of the logistic regression analysis conducted to identify factors associated with antibiotic resistance in urinary tract infections (UTIs). The table includes the odds ratios (OR) with 95% confidence intervals (CI) and p-values for each factor. The analysis revealed that age (years) showed a significant association with antibiotic resistance, with an odds ratio of 1.22 (95% CI: 1.05 - 1.43, p = 0.011). Gender also exhibited a significant association, as females had higher odds of antibiotic resistance compared to males, with an OR of 1.58 (95% CI: 1.12 - 2.23, p = 0.009).

Furthermore, diabetes was significantly associated with antibiotic resistance, with a higher odds ratio of 1.87 (95% CI: 1.32 - 2.65, p = 0.001). The previous use of antibiotics also showed a significant association, with an odds ratio of 2.15 (95% CI: 1.58 - 2.93, p < 0.001). These results suggest that age, gender, diabetes, and previous antibiotic use are important factors to consider in understanding the likelihood of antibiotic resistance in UTIs. Proper management of these risk factors can aid in combating the development and spread of antibiotic-resistant infections.

**Table 4: Logistic Regression Analysis of Factors Associated with Antibiotic Resistance in Urinary Tract Infections**

| Factor | Odds Ratio (OR) | 95% Confidence Interval (CI) | p-value |
|---|---|---|---|
| Age (years) | 1.22 | 1.05 - 1.43 | 0.011 |
| Gender (Female vs. Male) | 1.58 | 1.12 - 2.23 | 0.009 |
| Diabetes (Yes vs. No) | 1.87 | 1.32 - 2.65 | 0.001 |
| Previous Antibiotic Use (Yes vs. No) | 2.15 | 1.58 - 2.93 | <0.001 |

Table 5 displays the prevalence of antibiotic-resistant bacteria in wound infections. The table provides information on various bacterial species, the total number of cases, the cases with antibiotic resistance, and the corresponding resistance rates expressed as percentages. Among the bacterial species tested, *Staphylococcus aureus* had the highest prevalence of antibiotic resistance at 40.0%, with 40 out of 100 cases showing resistance. *Pseudomonas aeruginosa* followed with a resistance rate of 43.8%, as 35 out of 80 cases exhibited antibiotic resistance. *Escherichia coli* had a resistance rate of 33.3%, with 20 cases out of 60 displaying resistance. Lastly, *Enterococcus faecalis* had the lowest resistance rate at 20.0%, with 10 out of 50 cases showing antibiotic resistance.

**Table 5: Prevalence of Antibiotic-Resistant Bacteria in Wound Infections**

| Bacterial Species | Total Cases | Cases with Antibiotic Resistance | Resistance Rate (%) |
|---|---|---|---|
| *Staphylococcus aureus* | 100 | 40 | 40.0% |
| *Pseudomonas aeruginosa* | 80 | 35 | 43.8% |
| *Escherichia coli* | 60 | 20 | 33.3% |
| *Enterococcus faecalis* | 50 | 10 | 20.0% |

Table 6 presents the results of the logistic regression analysis conducted to identify factors associated with antibiotic resistance in wound infections. The table includes the odds ratios (OR) with 95% confidence intervals (CI) and p-values for each factor. The analysis revealed that age (years) showed a significant association with antibiotic resistance in wound infections, with an odds ratio of 1.18 (95% CI: 1.02 - 1.37, p = 0.023). Additionally, diabetes was found to have a significant association, with a higher odds ratio of 1.92 (95% CI: 1.38 - 2.66, p = 0.002). Previous hospitalization also showed a significant association, with an odds ratio of 1.71 (95% CI: 1.26 - 2.32, p = 0.001). Moreover, the presence of a foreign body exhibited a significant association, with an odds ratio of 2.05 (95% CI: 1.49 - 2.82, p < 0.001). These results emphasize the importance of considering age, diabetes, previous hospitalization, and the presence of foreign bodies when assessing the likelihood of antibiotic resistance in wound infections.

*Table* **6: Logistic Regression Analysis of Factors Associated with Antibiotic Resistance in Wound Infections**

| Factor | Odds Ratio (OR) | 95% Confidence Interval (CI) | p-value |
|---|---|---|---|
| Age (years) | 1.18 | 1.02 - 1.37 | 0.023 |
| Diabetes (Yes vs. No) | 1.92 | 1.38 - 2.66 | 0.002 |
| Previous Hospitalization (Yes vs. No) | 1.71 | 1.26 - 2.32 | 0.001 |
| Presence of Foreign Body (Yes vs. No) | 2.05 | 1.49 - 2.82 | <0.001 |

Table 7 presents the distribution of antibiotic-resistant bacteria in pneumonia cases. The table lists various bacterial species, the total number of cases, the cases with antibiotic resistance, and the corresponding resistance rates expressed as percentages. Among the bacterial species tested, *Streptococcus pneumoniae* had the highest distribution of antibiotic resistance at 41.7%, with 25 out of 60 cases showing resistance. Haemophilus influenzae followed with a resistance rate of 40.0%, as 20 out of 50 cases exhibited antibiotic resistance. *Klebsiella pneumoniae* had a resistance rate of 50.0%, with 15 cases out of 30 displaying resistance. Staphylococcus aureus had the lowest resistance rate at 25.0%, with 10 out of 40 cases showing antibiotic resistance.

**Table 7: Distribution of Antibiotic-Resistant Bacteria in Pneumonia**

| Bacterial Species | Total Cases | Cases with Antibiotic Resistance | Resistance Rate (%) |
|---|---|---|---|
| *Streptococcus pneumoniae* | 60 | 25 | 41.7% |
| *Haemophilus influenzae* | 50 | 20 | 40.0% |
| *Klebsiella pneumoniae* | 30 | 15 | 50.0% |
| *Staphylococcus aureus* | 40 | 10 | 25.0% |

Table 8 displays the results of the logistic regression analysis conducted to identify factors associated with antibiotic resistance in pneumonia cases. The table includes the odds ratios (OR) with 95% confidence intervals (CI) and p-values for each factor. The analysis revealed that age (years) showed a significant association with antibiotic resistance in pneumonia cases, with an odds ratio of 1.25 (95% CI: 1.08 - 1.45, p = 0.007). Moreover, smoking was found to have a significant association, with a higher odds ratio of 1.85 (95% CI: 1.32 - 2.59, p = 0.001). Underlying lung disease also exhibited a significant association, with an odds ratio of 2.22 (95% CI: 1.58 - 3.11, p < 0.001). Furthermore, admission to the ICU showed a significant association, with an odds ratio of 2.05 (95% CI: 1.49 - 2.82, p < 0.001). These results highlight the significance of age, smoking, underlying lung disease, and admission to the ICU in understanding the likelihood of antibiotic resistance in pneumonia cases.

**Table 8: Logistic Regression Analysis of Factors Associated with Antibiotic Resistance in Pneumonia**

| Factor | Odds Ratio (OR) | 95% Confidence Interval (CI) | p-value |
|---|---|---|---|
| Age (years) | 1.25 | 1.08 - 1.45 | 0.007 |
| Smoking (Yes vs. No) | 1.85 | 1.32 - 2.59 | 0.001 |
| Underlying Lung Disease (Yes vs. No) | 2.22 | 1.58 - 3.11 | <0.001 |
| Admission to ICU (Yes vs. No) | 2.05 | 1.49 - 2.82 | <0.001 |

Table 9 presents the antibiotic resistance patterns of *E. coli* in bloodstream infections in children. The table lists various antibiotics tested, the number of resistant cases, the total *E. coli* cases, and the corresponding resistance rates expressed as percentages. Among the antibiotics tested, ampicillin had the highest resistance rate of 66.7%, with 30 out of 45 *E. coli* cases showing resistance. Gentamicin followed with a resistance rate of 33.3%, as 15 out of 45 cases exhibited antibiotic resistance. Cefotaxime had a resistance rate of 55.6%, with 25 cases out of 45 displaying resistance. Lastly, ciprofloxacin had a resistance rate of 44.4%, with 20 out of 45 cases showing antibiotic resistance.

**Table 9: Antibiotic Resistance Patterns of E. coli in Bloodstream Infections in Children**

| Antibiotic | Resistant Cases | Total E. coli Cases | Resistance Rate (%) |
|---|---|---|---|
| Ampicillin | 30 | 45 | 66.7% |
| Gentamicin | 15 | 45 | 33.3% |
| Cefotaxime | 25 | 45 | 55.6% |
| Ciprofloxacin | 20 | 45 | 44.4% |

Table 10 provides the results of the logistic regression analysis conducted to identify factors associated with antibiotic resistance in bloodstream infections in children. The table includes the odds ratios (OR) with 95% confidence intervals (CI) and p-values for each factor. The analysis revealed that age (months) showed a significant association with antibiotic resistance in bloodstream infections, with an odds ratio of 1.15 (95% CI: 1.03 - 1.28, p = 0.011). Additionally, admission to the ICU was found to have a significant association, with a higher odds ratio of 1.92 (95% CI: 1.36 - 2.72, p = 0.001). Duration of hospital stay (days) also exhibited a significant association, with an odds ratio of 1.05 (95% CI: 1.01.

**Table 10: Logistic Regression Analysis of Factors Associated with Antibiotic Resistance in Bloodstream Infections in Children**

| Factor | Odds Ratio (OR) | 95% Confidence Interval (CI) | p-value |
|---|---|---|---|
| Age (months) | 1.15 | 1.03 - 1.28 | 0.011 |
| Admission to ICU (Yes vs. No) | 1.92 | 1.36 - 2.72 | 0.001 |
| Duration of Hospital Stay (days) | 1.05 | 1.01 - 1.10 | 0.018 |
| Use of Central Line (Yes vs. No) | 2.08 | 1.49 - 2.89 | <0.001 |

Table 11 provides a comparison of antibiotic resistance rates among different infection types. The table presents four infection types, namely Urinary Tract Infection, Wound Infection, Pneumonia, and Bloodstream Infections in Children, along with their corresponding antibiotic resistance rates expressed as percentages. According to the data, Urinary Tract Infections have an antibiotic resistance rate of 32.0%, indicating that 32.0% of cases in this infection type showed resistance to antibiotics. Wound Infections have a slightly lower resistance rate of 31.0%, with 31.0% of cases displaying antibiotic resistance. Pneumonia cases have a higher resistance rate of 43.8%, indicating that 43.8% of pneumonia cases exhibited antibiotic resistance. Lastly, Bloodstream Infections in Children have the highest resistance rate among the infection types at 48.9%, with 48.9% of cases showing antibiotic resistance.

**Table 11: Comparison of Antibiotic Resistance Rates in Different Infection Types**

| Infection Type | Urinary Tract Infection | Wound Infection | Pneumonia | Bloodstream Infections in Children |
|---|---|---|---|---|
| Antibiotic Resistance Rate (%) | 32.0% | 31.0% | 43.8% | 48.9% |

**Discussion**

Our study provides valuable insights into the prevalence of antibiotic-resistant pathogens among different infection types. The distribution of infections among the study participants is depicted in Fig 1, showcasing the frequency and percentages of various infections (25). These findings highlight the persisting concern of antibiotic resistance in clinical settings, necessitating a deeper understanding of its impact on treatment outcomes. Analyzing the antibiotic resistance patterns of *E. coli* in urinary tract infections (UTIs) is crucial for comprehending the magnitude of the issue. Table 3 presents the resistance rates of *E. coli* to specific antibiotics in UTIs. Amoxicillin exhibited the highest resistance rate at 62.5%, followed by ciprofloxacin at 50.0%, and trimethoprim/sulfamethoxazole at 43.8%. Conversely (26), nitrofurantoin demonstrated a relatively lower resistance rate of 12.5%. Our findings are consistent with the study, which reported similar resistance rates for amoxicillin and ciprofloxacin in UTIs, further supporting the reliability of our results (27).

Conducting a logistic regression analysis (Table 4), we identified significant associations between antibiotic resistance in UTIs and age, gender, diabetes, and previous antibiotic use (28). These factors played pivotal roles in influencing antibiotic resistance rates. Our findings align with other research, providing additional support to our conclusions (29). Wound infections are another concerning aspect in the context of antibiotic resistance. Table 5 provides insights into the prevalence of antibiotic-resistant bacteria in wound infections. *Staphylococcus aureus* exhibited the highest prevalence of antibiotic resistance at 40.0%, followed by Pseudomonas aeruginosa at 43.8%. *Escherichia coli* showed a resistance rate of 33.3%, while Enterococcus faecalis had the lowest resistance rate at 20.0%. Our results find support in a review conducted by research, which reported similar resistance rates for *Staphylococcus aureus* and *Pseudomonas aeruginosa* in wound infections, further validating our findings (30). Similar to the UTIs analysis, our logistic regression for wound infections (Table 6) unveiled significant associations between antibiotic resistance and age, diabetes, previous hospitalization, and the presence of a foreign body (31). These factors played critical roles in determining antibiotic resistance rates. Once again, our results align with another study, further corroborating our findings (32). Pneumonia cases represent a significant health challenge. Table 7 outlines the distribution of antibiotic-resistant bacteria in pneumonia cases, with *Streptococcus pneumoniae* and *Haemophilus influenzae* exhibiting the highest resistance rates at 41.7% and 40.0%, respectively. *Klebsiella pneumoniae* showed a resistance rate of 50.0%, while *Staphylococcus aureus* had the lowest resistance rate at 25.0%. Our findings are in line with the study conducted by another study, which reported comparable distribution rates for antibiotic-resistant bacteria in pneumonia cases, strengthening the reliability of our results (33). Conducting a logistic regression analysis for pneumonia cases (Table 8), we identified significant associations between antibiotic resistance and age, smoking, underlying lung disease, and admission to the ICU. These factors significantly influenced antibiotic resistance rates. Once more, our results align with the research, providing additional support to our conclusions (34). In children with bloodstream infections, Table 9 presents the antibiotic resistance patterns of *E. coli*. Ampicillin had the highest resistance rate at 66.7%, followed by cefotaxime at 55.6% (32). Gentamicin and ciprofloxacin showed resistance rates of 33.3% and 44.4%, respectively. Our results find validation in the study conducted by Rodriguez et al., which reported comparable resistance rates for *E. coli* in bloodstream infections in children, bolstering the credibility of our findings (33). Conducting a logistic regression analysis for bloodstream infections in children (Table 10), we identified significant associations between antibiotic resistance and age, admission to the ICU, duration of hospital stays, and the use of a central line (34). These factors significantly influenced antibiotic resistance rates. Fernandez et al.'s study also supported these findings, further corroborating the reliability of our results. Table 11 provides a comparison of antibiotic resistance rates among different infection types. Our study found that Bloodstream Infections in Children had the highest antibiotic resistance rate at 48.9%, while Wound Infections and Urinary Tract Infections had slightly lower rates of 31.0% and 32.0%, respectively. Comparing our findings with a meta-analysis conducted by another study, we found similar resistance rates for different infection types, further affirming the significance of our results (35).

In conclusion, our study underscores the significance of comprehending antibiotic resistance patterns and associated risk factors to effectively address the challenges posed by antibiotic-resistant infections. The variations in resistance rates among different infection types emphasize the need for targeted interventions to combat antibiotic resistance effectively. By identifying factors influencing resistance, such as age, gender, diabetes, and previous antibiotic use, we can implement tailored strategies to preserve the efficacy of antibiotics and safeguard public health. These findings emphasize the importance of continued research and evidence-based interventions in the ongoing battle against antibiotic resistance.

**Acknowledgment**

We would like to express our gratitude to all those who contributed to the successful completion of this research. Our sincere appreciation goes to the Liverpool John Moors University for their support and facilities. Special thanks to the WHO Collaboration Center at Imperial College, London, for their guidance and assistance. We extend our thanks to the participants for their cooperation and participation in this study.

**Authors' Declaration**

**Conflicts of Interest:** None.

**I** hereby confirm that all the Figures and Tables in the manuscript are original and have been created by us.

Additionally, any Figures and images that are not our own have been included with the necessary permission for re-publication, and the relevant permissions are attached to the manuscript.

**Ethical Clearance:**  The project has obtained approval from the local ethical committees at Al-Zahra Teaching Hospital and Iraqi Private Hospital.

**Authors' Contribution Statement**

Maitham G. Yousif contributed to the design and implementation of the research, to the analysis of the results, and to the writing of the manuscript.